\documentclass[12pt]{article}
\usepackage{graphicx}
\newcommand\pubnumber{IPPP/23/33}
\newcommand\pubdate{\today}

\def\Title#1{\begin{center} {\Large #1 } \end{center}}
\def\Author#1{\begin{center}{ \sc #1} \end{center}}
\def\Address#1{\begin{center}{ \it #1} \end{center}}

\newcommand\pubblock{\rightline{\begin{tabular}{l} \pubnumber\\
         \pubdate  \end{tabular}}}
\newenvironment{Abstract}{\begin{quotation}  }{\end{quotation}}
\newenvironment{Presented}{\begin{quotation} \begin{center} 
             PRESENTED AT\end{center}\bigskip 
      \begin{center}\begin{large}}{\end{large}\end{center} \end{quotation}}

\usepackage{amsmath}
\usepackage{graphicx}
\usepackage[backend=biber,style=numeric-comp,
doi=false,isbn=false,url=false,eprint=false,
sorting=none]{biblatex}
\usepackage{xspace}

\newcommand\lep{L\scalebox{0.8}{EP}\xspace}

\newcommand\hera{H\scalebox{0.8}{ERA}\xspace}

\newcommand{\MSbar}{\ensuremath{\overline{\mathrm{MS}}}}

\newcommand{\NLO}{\ensuremath{\mathrm{NLO}}\xspace}

\newcommand{\sla}[1]{\ensuremath{{#1\kern-0.45em/}}}

\newcommand{\MCatNLO}{M\protect\scalebox{0.8}{C}@N\protect\scalebox{0.8}{LO}\xspace}

\newcommand{\Sherpa}{S\protect\scalebox{0.8}{HERPA}\xspace}

\begin{document}
\begin{titlepage}
 \pubblock
\vfill
\Title{Full event simulation of Photoproduction at \NLO QCD in \Sherpa}
\vfill
\Author{Peter Meinzinger}
\Address{
  Institute for Particle Physics Phenomenology, \\
  Durham University, Durham DH1 3LE, UK
}
\vfill
\begin{Abstract}
\noindent
Photoproduction is an important mode for the production of jets and electro-weak particles at lepton--lepton and lepton--hadron colliders and allows for interesting studies of exclusive production at hadron--hadron colliders. In this talk, I will review recent efforts of extending the \Sherpa event generator to include the calculation of photoproduction cross sections for electron and proton beams, including the simulation of underlying events. The framework is validated using data of jet production at the \hera and \lep experiments and lepton production at the LHC. I will discuss advances towards achieving matched \NLO accuracy and fully capturing the dynamics of inclusive and exclusive photoproduction at different colliders.
\end{Abstract}
\vfill
\begin{Presented}
DIS2023: XXX International Workshop on Deep-Inelastic Scattering and
Related Subjects, \\
Michigan State University, USA, 27-31 March 2023 \\
     \includegraphics[width=9cm]{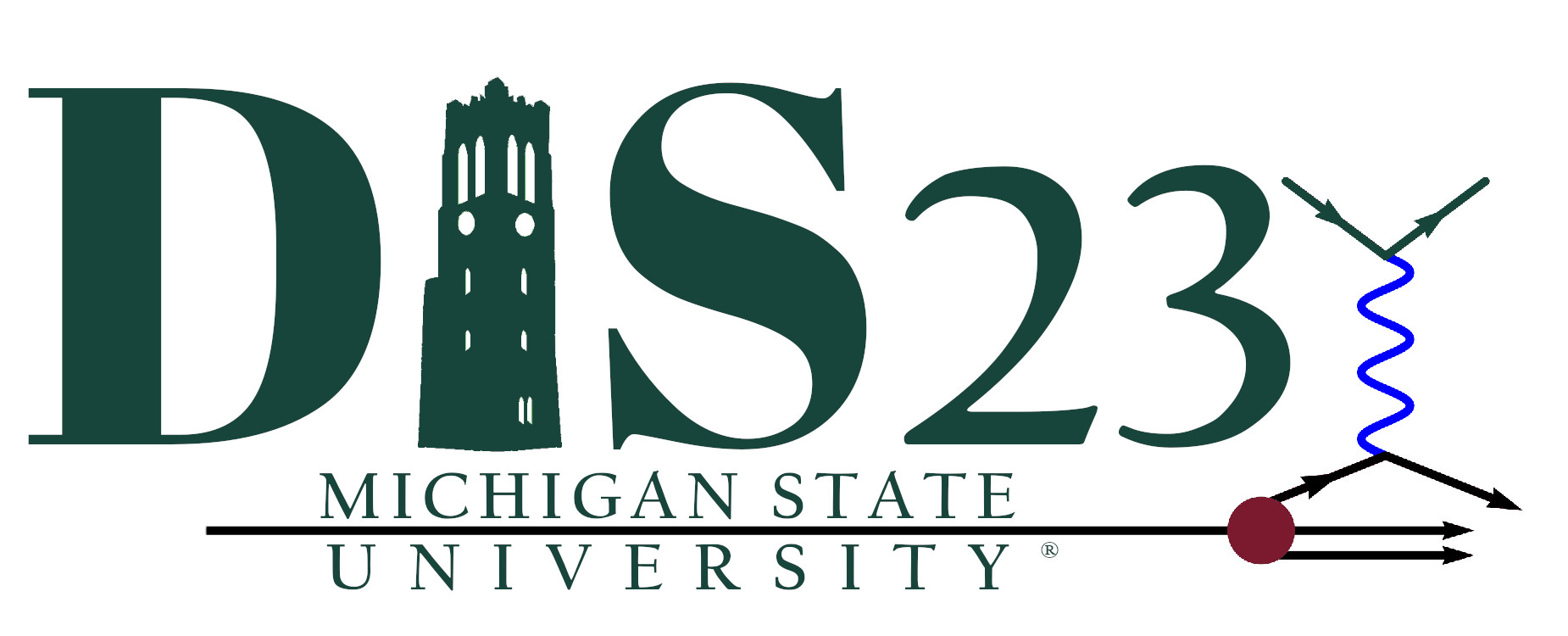}
\end{Presented}
\vfill
\end{titlepage}

\section{Introduction}

The cross section of jet production at lepton--hadron or lepton--lepton collider experiments is dominated by the exchange of a virtual photon. While, in particular at the latter, this is well understood at large photon virtualities, the descriptive power of the theoretical calculations deteriorates with decreasing virtuality~\cite{Carli:2010jb}. This has been reflected in decomposing the full cross section into electro- and photoproduction where the latter is identified with a regime where the photon is quasi-real and has to be seen as the incoming particle. 
Simulating these events needs a different approach than the typical DIS processes. Here we report on the implementation of relevant physics and its validatation in \Sherpa. 

\section{Simulation in \Sherpa}

\subsection{Photon flux}

As the electron decouples from the hard interaction in the scattering, the flux of the quasi-real photons has to be calculated. In the  Weizsäcker-Williams approximation~\cite{Budnev:1975poe} the cross section is calculated as 

\begin{equation}
    \mathrm{d}\sigma_{e p \to e^\prime + 2j + X} = \sigma_{\gamma p \to 2j + X}(x, s)|_{Q^2=0} \ \mathrm{d}n(x)\,,
\end{equation}
where the electron momentum can be reconstructed from the photon and the photon virtuality is integrated out in the equivalent photon flux $\mathrm{d}n$, leaving only the maximum virtuality $Q^2_\mathrm{max}$ as a free parameter, which has to be determined by the experimental setup and by the considered process. 


For the measurements considered in this study, the photon flux for electron beams includes a mass-dependent correction as proposed in~\cite{Frixione:1993yw}: 
\begin{equation}
    \mathrm{d}n(x) = \frac{\alpha_\mathrm{em}}{2 \pi} \frac{\mathrm{d}x}{x}
			\left[ \left[ 1 + (1 - x)^2 \right] \log \left( \frac{Q^2_\mathrm{max}}{Q^2_\mathrm{min}} \right) +
			2 m_e^2 x^2 \left( \frac{1}{Q^2_\mathrm{min}} - \frac{1}{Q^2_\mathrm{max}} \right) \right]
\end{equation}
Here, $x$ is the fraction of the photon momentum with respect to the electron momentum, $m_e$ is the electron mass and $Q_\mathrm{min/max}$ are the minimum and maximum photon virtualities, where the former is given by kinematic constraints as $Q^2_\mathrm{min} = \frac{m_e^2 x^2}{1 - x}$.

\subsection{Parton distributions in the photon}
Initial State Radiation off the photon cannot be neglected in photoproduction of jets, necessitating the inclusion of the resolved photon component in the calculation, i.e.\ its hadronic structure. Hence~\cite{Frixione:1993yw}: 

\begin{align}
    \mathrm{d}\sigma_{\gamma p \to 2 j + X} &= \mathrm{d}\sigma_{\gamma p \to 2 j + X}^\mathrm{(hl)} + \mathrm{d}\sigma_{\gamma p \to 2 j + X}^\mathrm{(pl)} \ \mathrm{, with}\nonumber\\
    \mathrm{d}\sigma_{\gamma p \to 2 j + X}^\mathrm{(hl)} &= \sum_{ij} \int \mathrm{d}x f_{i/\gamma}(x, \mu_F^\prime) f_{j/p}(x, \mu_F) \mathrm{d}\hat{\sigma}_{ij} (p_\gamma, x p_p, \alpha_S, \mu_R, \mu_F, \mu_F^\prime) \nonumber\\
    \mathrm{d}\sigma_{\gamma p \to 2 j + X}^\mathrm{(pl)} &= \sum_j \int \mathrm{d}x f_{j/p}(x, \mu_F) \mathrm{d}\hat{\sigma}_{\gamma j} (p_\gamma, x p_p, \alpha_S, \mu_R, \mu_F, \mu_F^\prime)\,,
\end{align}
where the superscripts stand for the hadron- and point-like photon respectively, the $f_{i/A}$ are the parton distribution functions (PDFs) related to finding parton $i$ in particle $A$, the $\mu_{F,\,R}$ are the factorisation and renormalisation scales, and $p$ are the momenta. 

The photon PDF obeys an evolution slightly different to hadronic PDFs, due to the presence of a QED splitting kernel, leading to

\begin{equation}
    \frac{\partial f_{i/\gamma}}{\partial \mathrm{log} \mu^2} = \frac{\alpha_S}{2 \pi} \sum_j P_{ij} \otimes f_{j/\gamma} + \frac{\alpha_\mathrm{em}}{2 \pi} P_{i\gamma}
\end{equation}
with $P$ the splitting kernels and where the first term is the usual QCD evolution and the latter the QED evolution stemming from a photon splitting into two quarks. 

\begin{figure}
    \centering
    \begin{minipage}[t]{0.49\textwidth}
        \centering
        \includegraphics[trim={0 0 0 14pt}, width=.9\textwidth,clip]{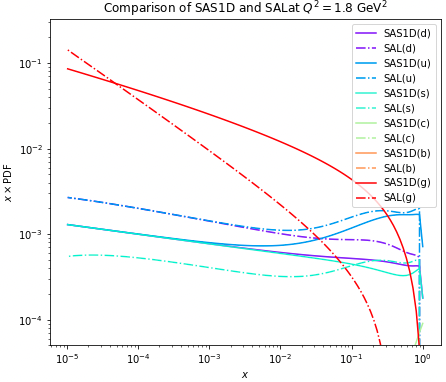}
        \caption{Comparison of the SAS1D and the SAL pdf sets at $\mu^2 = 1.8 \ \mathrm{GeV}^2$. }
        \label{fig.pdfs-comparison}
    \end{minipage}
    \hfill
    \begin{minipage}[t]{0.49\textwidth}
        \centering
        \includegraphics[width=.9\linewidth]{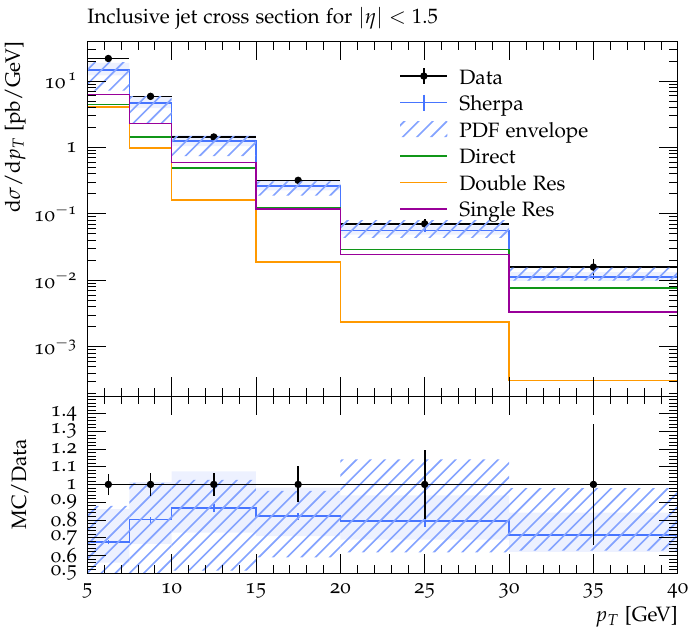}
        \caption{Distribution for jet transverse momentum $p_T$ for \lep at $\sqrt{s} = 206$ GeV, averaged over all 10 PDF sets. }
        \label{fig.lo-study}
    \end{minipage}
\end{figure}

Photon PDFs from Glück-Reya-Vogt \cite{Gluck:1991jc}, Glück-Reya-Schienbein \cite{Gluck:1999ub}, Slominski-Abramowicz-Levy \cite{Slominski:2005bw}, and Schuler-Sjöstrand \cite{Schuler:1995fk,Schuler:1996fc} have been included in \Sherpa. 
As exemplified in a comparison between two PDF sets (the SAS1D paramterisation by Schuler-Sjöstrand and the set by Slominski-Abramowicz-Levy) in Fig.\ \ref{fig.pdfs-comparison}, there are large deviations, especially in the gluon distribution function. 

The distinction of direct and resolved processes can not be maintained at Next-to-Leading-Order (NLO) due to the ambiguity of real emissions. While the resolved-photon processes can be computed at \NLO analogously to jet production in $p$-$p$ collisions, the direct-photon processes show divergences stemming from the photon splittings $P_{i\gamma}$. However, in~\cite{Frixione:1993dg} it was shown that these divergences cancel against the resolved-photon cross-section as these splittings are re-absorbed into the PDF by means of the inhomogenous term proportional to $P_{i\gamma}$ in the evolution equation. Hence, these divergences can be subtracted from $\mathrm{d}\sigma_{\gamma p \to 2 j + X}^\mathrm{(pl)}$ and care only has to be taken to use a photon PDF with the correct evolution and the same factorisation scheme as in the matrix element generation. The calculation can then be matched to the parton shower with the \MCatNLO prescription. The main difference lies in the fact that momentum fractions have to be calculated with respect to the variable photon energies instead of fixed beam energies. 

\section{Validation}

For validation, the \Sherpa simulation has been compared to data from the \lep and \hera colliders, namely photoproduction of one or two jets at the ZEUS, OPAL and L3 experiments. Typical observables in these analyses are the (average) jet transverse energy $E_T$, pseudo-rapidity $\eta$, $\cos \Theta^*$, which approximates the angle between two jets and $x_\gamma^\pm$, which is defined as

\begin{equation}
    x_\gamma^\pm = ( \sum\limits_{j=1,2} E^{(j)}\pm p_z^{(j)}) / ( \sum\limits_{i\in{\rm hfs}} E^{(i)}\pm p_z^{(i)} )
\end{equation}
and works as a proxy to experimentally distinguish the direct from the resolved modes. 

\begin{figure}
    \centering
    \begin{minipage}[t]{0.49\textwidth}
        \centering
        \includegraphics[width=.9\linewidth]{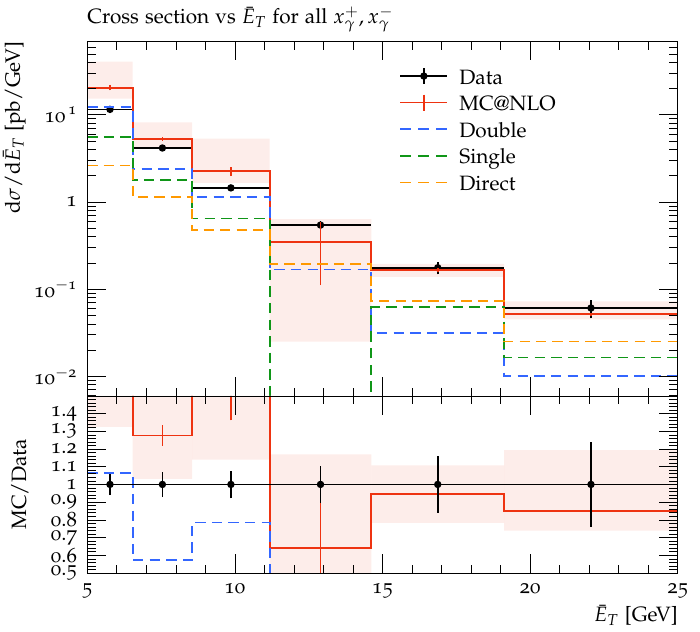}
        \caption{Average jet transverse energy $\bar{E}_T$ for \lep at $\sqrt{s} = 198$ GeV. }
        \label{fig.nlo-lep}
    \end{minipage}
    \hfill
    \begin{minipage}[t]{0.49\textwidth}
        \centering
        \includegraphics[width=.9\linewidth]{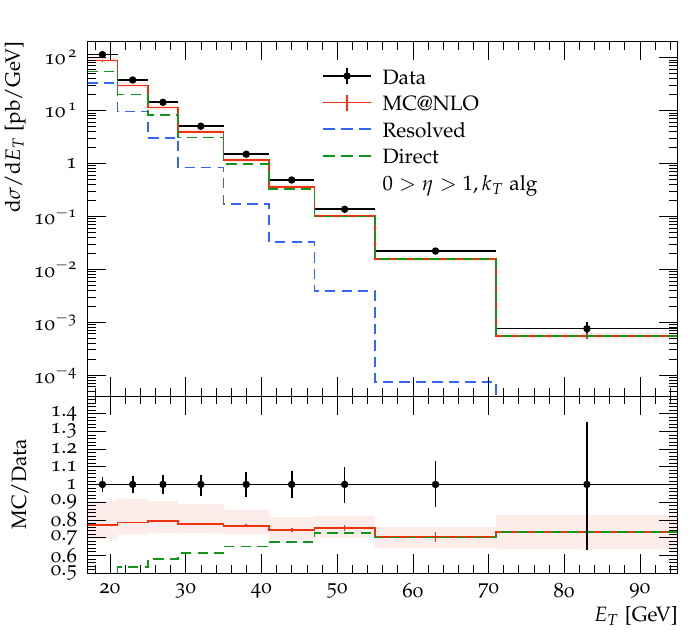}
        \caption{Distribution for jet transverse energy $E_T$ for \hera. }
        \label{fig.nlo-hera}
    \end{minipage}
\end{figure}

In Fig.\ \ref{fig.lo-study} we studied at LO, where all PDF sets could be used, the uncertainties from the different PDF parametrisations and found significant deviations, in agreement with the large discrepancies in the parton distributions. This underlines the need for a new fit to the available data and a more thorough study of the parton distribution of the real and quasi-real photon. Overall, the simulation shows good agreement with the data within the uncertainties. The results at \NLO, cf.\ Figs.\ \ref{fig.nlo-lep} and \ref{fig.nlo-hera}, were generated as an average over the SAS1M and SAS2M pdf sets, which use the \MSbar~scheme. 

\section{Outlook}

\subsection{Minimum Bias photoproduction for the LHC}

Multiple-parton interactions are non-negligible in photoproduction~\cite{Butterworth:1996zw} and the implementation based on~\cite{Sjostrand:1987su} has been extended to also cover parametrisations of $\gamma p$ and $\gamma \gamma$ interactions. 
One object of study could be the simulation of Minimum Bias events where interactions are not only allowed between the two proton beams, but also the photon--proton and photon--photon systems to examine systems with rapidity gaps at the LHC.
When studying semi-diffractive processes, e.g. at the LHC, the LUXqed PDF can be used to access both the elastic and the dissociative contributions to the photoproduction processes. 

\subsection{Diffractive photoproduction and pomeron exchange}

The diffractive production of jets is often understood in terms of a pomeron exchange which is factorized into a pomeron flux and a pomeron parton distribution. At \hera the factorisation was observed to break down, so there's ongoing interest to understand this phenomenon~\cite{Guzey:2016awf}, especially in view of the upcoming Electron-Ion Collider. 
The implementation of the pomeron flux is work in progress in \Sherpa. 

\section{Summary}

We showed progress in \Sherpa to include photoproduction at various colliders and achieving matched \NLO accuracy in QCD. The validation has been done at LO and is ongoing for \NLO. We also discussed several ideas how to extend the framework further and use it in experimental studies at the LHC and the EIC. 

\printbibliography
 
\end{document}